\documentstyle[12pt,epsf]{article}
\def \beq{\begin{equation}}
\def \eeq{\end{equation}}
\def \s{\sqrt{2}}
\def \st{\sqrt{3}}
\def \sx{\sqrt{6}}
\begin{document}
\renewcommand{\thetable}{\Roman{table}}
\rightline{TECHNION-PH-95-26}
\rightline{EFI-95-59}
\rightline{hep-ph/9509325}
\rightline{September 1995}
\bigskip
\bigskip
\centerline{\bf DETERMINING THE WEAK PHASE $\gamma$}
\centerline{\bf FROM CHARGED B DECAYS}
\bigskip
\centerline{\it Michael Gronau}
\centerline{\it Department of Physics}
\centerline{\it Technion -- Israel Institute of Technology, Haifa 32000,
Israel}
\medskip
\centerline{and}
\medskip
\centerline{\it Jonathan L. Rosner}
\centerline{\it Enrico Fermi Institute and Department of Physics}
\centerline{\it University of Chicago, Chicago, IL 60637}
\bigskip
\centerline{\bf ABSTRACT}
\medskip
\begin{quote}

A quadrangle relation is shown to be satisfied by the amplitudes for $B^+ \to
\pi^0 K^+,~\pi^+ K^0,~\eta K^+$, and $\eta' K^+$.  By comparison with the
corresponding relation satisfied by $B^-$ decay amplitudes, it is shown that
the relative phases of all the amplitudes can be determined up to discrete
ambiguities.  Making use of an SU(3) relation between amplitudes contributing
to the above decays and those contributing to $B^{\pm} \to \pi^{\pm} \pi^0$, it
is then shown that one can determine the weak phase $\gamma \equiv {\rm Arg}
(V_{ub}^* V_{cb}/V_{us}^* V_{cs})$, where $V$ is the Cabibbo-Kobayashi-Maskawa
matrix describing the charge-changing weak interactions between the quarks
$(u,c,t)$ and $(d,s,b)$.

\end{quote}
\newpage

\centerline{\bf I. INTRODUCTION}
\bigskip

The presence of phases in elements of the Cabibbo-Kobayashi-Maskawa (CKM)
matrix \cite{CKM} is the currently favored explanation of the observed
violation of CP invariance \cite{CCFT} in the neutral kaon system.  However,
independent tests for the presence of these phases are needed.  One class of
such tests is based on observing the predicted violation of CP invariance in
$B$ meson decays.  In order to interpret such violations in terms of CKM phases
one needs to identify the flavor of the decaying $B$ meson, posing potential
``tagging'' problems for neutral $B$'s, or to understand final-state
interactions in ``self-tagging'' modes such as $B^{\pm} \to \pi K$.

Using flavor SU(3) [3--6] we proposed [7--9] that $B^\pm \to \pi \pi$ and
$B^\pm \to \pi K$ amplitudes could be related to one another in such a way as
to obtain the weak phase $\gamma \equiv {\rm Arg} (V_{ub}^* V_{cb}/V_{us}^*
V_{cs})$, where $V_{ij}$ is the element of the CKM matrix describing the
charge-changing weak interaction between the quarks $i = (u,c,t)$ and $j =
(d,s,b)$.  It was shown \cite{DH,DHT} that electroweak penguin contributions
\cite{RF} could spoil this relation, and were likely to be significant.  We
proposed a way to specify the phase $\gamma$ independently of these
contributions but making use of a rare decay $B_s \to \pi^0 \eta$ \cite{EWP}.
A simpler relation yields similar information by employing the decays $B^\pm
\to \eta_8 K^\pm$, where $\eta_8$ denotes an octet member.  The amplitude
triangle relation \cite{Desh}
\beq \label{eqn:DH}
A(B^+ \to \pi^0 K^+) + \sqrt{2} A(B^+ \to \pi^+ K^0) = \sqrt{3}
A(B^+ \to \eta_8 K^+)~~~
\eeq
(also implied by the first three lines of Table II of Ref.~\cite{BPP}) can be
compared with the relation for the charge-conjugate process to yield an
amplitude which, when normalized by that governing $B^\pm \to \pi^\pm \pi^0$,
yields the phase $\gamma$.

A modified version of the triangle relation (\ref{eqn:DH}) can be applied to
the physical $\eta$, an octet-singlet mixture, only if one neglects the
independent amplitudes in which the singlet component of the $\eta$ is
produced.  In the present paper we identify the most important of these
amplitudes and show how to recover information on $\gamma$ in its presence.

We shall show that the natural generalization of (\ref{eqn:DH}) for an
octet-singlet mixture, the quadrangle relation
\beq \label{eqn:firstquad}
A(B^+ \to \pi^0 K^+) + \sqrt{2} A(B^+ \to \pi^+ K^0) = \sqrt{3}
[\cos \theta A(B^+ \to \eta K^+) + \sin \theta A(B^+ \to \eta' K^+)]~~~,
\eeq
where
\beq \label{eqn:etadefs}
\eta \equiv \eta_8 \cos \theta - \eta_1 \sin \theta~~~,~~
\eta' \equiv \eta_8 \sin \theta + \eta_1 \cos \theta~~~,
\eeq
and
$\eta_8 \equiv (2 s \bar s - u \bar u - d \bar d)/\sqrt{6}$,
$\eta_1 \equiv (s \bar s + u \bar u + d \bar d)/\sqrt{3}$,
can be compared with the corresponding relation for $B^-$ decays in order to
learn the shape of both quadrangles.  One can then form a difference between
two amplitudes for $B \to \pi K$ and $\bar B \to \pi \bar K$ which, when
compared with the amplitude for $B^\pm \to \pi^\pm \pi^0$, provides the weak
phase $\gamma$.

We introduce notation and assumptions and describe decay amplitudes in terms of
four independent quantities in Sec.~II.  The quadrangle (\ref{eqn:firstquad})
and that for $B^-$ decays are constructed in Sec.~III, where we also obtain an
expression for $\gamma$.  Some comments regarding SU(3) breaking and
experimental considerations occupy Sec.~IV, while Sec.~V concludes.  An
explicit geometric construction of amplitude quadrangles is described in an
Appendix.
\bigskip

\centerline{\bf II.  NOTATION, ASSUMPTIONS AND AMPLITUDES}
\bigskip

\leftline{\bf A. Definition of states}
\bigskip

For SU(3) amplitudes, we adopt a graphical notation described in more
detail in Refs.~\cite{BPP,PRL,PLB,EWP,SU3}.  Our states are defined by
$$
\pi^+ = u \bar d~~~,~~\pi^0 = (d \bar d - u \bar u)/\sqrt{2}~~~,~~
\pi^- = - d \bar u~~~,
$$
$$
K^+ = u \bar s~~~,~~K^0 = d \bar s~~~,~~\bar K^0 =  s \bar d~~~,~~
K^- = - s \bar u~~~,
$$
\beq
\eta_8 \equiv (2 s \bar s - u \bar u - d \bar d)/\sqrt{6}~~~,~~
\eta_1 \equiv (s \bar s + u \bar u + d \bar d)/\sqrt{3}~~~,
\eeq
with (\ref{eqn:etadefs}) describing the physical $\eta$ and $\eta'$. A good
approximation, corresponding to an octet-singlet mixing angle \cite{Chau,GK}
$\theta = \theta_p \equiv {\rm arcsin}~(1/3) \simeq 19.5^{\circ}$, is the
representation
\beq \label{eqn:mix}
\eta = \eta_p \equiv \frac{s \bar s - u \bar u - d \bar d}{\sqrt{3}}~~~,
{}~~\eta' = \eta'_p \equiv \frac{2 s \bar s + u \bar u + d \bar d}
{\sqrt{6}}~~~.
\eeq
\bigskip

\leftline{\bf B.  Assumptions about amplitudes}
\bigskip

The decays $B \to M_8 M_8$, where $M_8$ are pseudoscalar mesons belonging to
octets of flavor SU(3), are characterized by 5 independent amplitudes,
corresponding to one ${\bf 27}$, three ${\bf 8}$'s, and one ${\bf 1}$ in the
direct channel \cite{DZ}.  (We denote a flavor SU(3) representation by its
dimension in bold face.)  In previous works [7--9, 13, 15] we have argued that
the neglect of amplitudes containing factors of $f_B/m_B$ is equivalent to
relations between the ${\bf 27}$ and one of the ${\bf 8}$'s, and between the
${\bf 1}$ and another of the ${\bf 8}$'s, leaving 3 independent amplitudes.
These can be characterized by graphs $T,~C,~P$, illustrated in Fig.~1, in which
the spectator quark in the decaying $B$ does not enter into the decay
Hamiltonian.  A number of tests were proposed \cite{BPP} for the description of
$B$ decays in terms of this restricted set of SU(3) amplitudes.

\begin{figure}
\centerline{\epsfysize = 1.8in \epsffile{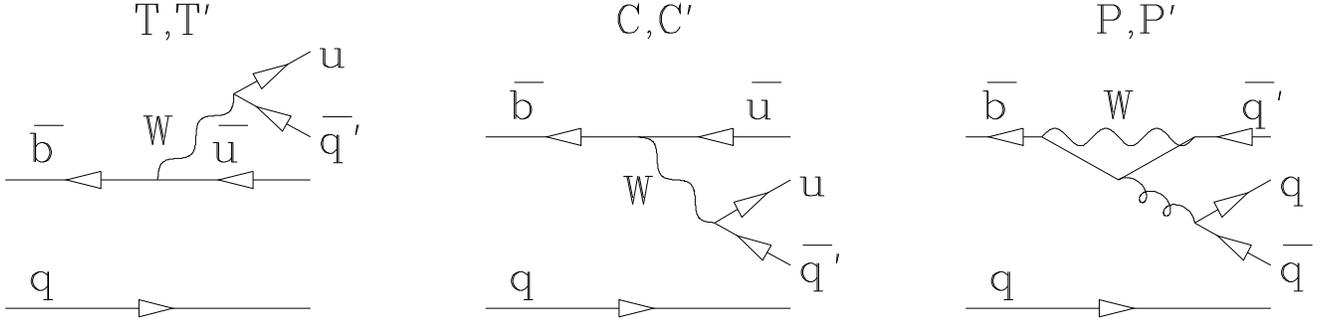}}
\caption{Graphs contributing to $B \to M_8 M_8$ decays which are not suppressed
by a factor of $f_B/m_B$.  Here $q = u,d,s$, while $q' = d$ for unprimed
amplitudes and $s$ for primed amplitudes.  The coiled line in the third graph
denotes exchange of one or more gluons.}
\end{figure}

The presence of electroweak penguin contributions [10--13] does not alter the
validity of an SU(3) description, as long as one relates amplitudes with the
same strangeness change ($|\Delta S| = 0$ or 1) to one another.  In that case
one may simply substitute
\beq
T \to t \equiv T + P_{EW}^C~~~,
\eeq
\beq
C \to c \equiv C + P_{EW}~~~,
\eeq
\beq
P \to p \equiv P - \frac{1}{3}P_{EW}^C~~~,
\eeq
where $P_{EW}$ and $P_{EW}^C$, the color-favored and color-suppressed
electroweak penguin amplitudes, correspond to the graphs in Fig.~2.

\begin{figure}
\centerline{\epsfysize = 1.8in \epsffile{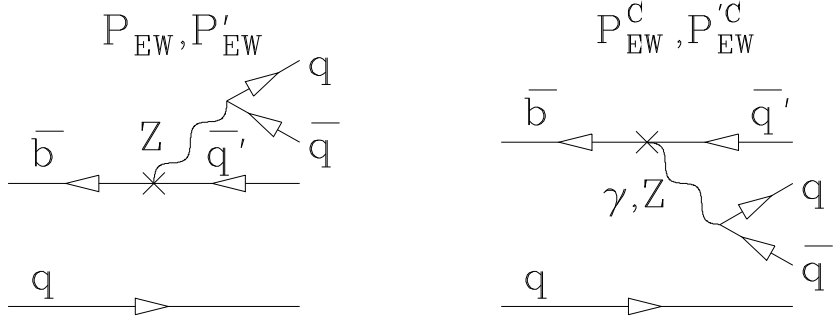}}
\caption{Electroweak penguin graphs contributing to $B \to M_8 M_8$ decays.
The crosses denote loops involving $W$ exchange.  Not shown are additional $WW$
box diagrams required for gauge invariance.}
\end{figure}

When one or two singlet pseudoscalar mesons $M_1$ are in the final state,
additional amplitudes must be taken into account \cite{DZ, Dighe}.  For the
decays $B \to M_1 M_8$, since the final state is an octet, there are three
${\bf 8}$ amplitudes corresponding to the three different representations in
the weak Hamiltonian $H_W$ describing $\bar b \to \bar q_1 q_2 \bar q_3$, where
$q_i$ are light $(u,d,s)$ quarks.  These representations transform as ${\bf
3^*},~{\bf 6}$, or ${\bf 15^*}$ of SU(3).  When combined with the ${\bf 3}$ of
the spectator quark, each contains one octet.  For the decays $B \to M_1 M_1$,
there is one singlet obtained from the product of the ${\bf 3^*}$ in $H_W$ and
the ${\bf 3}$ of the spectator quark.

As in the case of $M_8 M_8$ production, we now neglect all amplitudes in which
the spectator quark enters into the decay Hamiltonian. We thus identify a
single new amplitude, depicted in Fig.~3, contributing to $M_1 M_8$ and $M_1
M_1$ production.  This amplitude is denoted by $P_1$. This assumption has also
been adopted in Ref.~\cite{Dighe}, where many tests of it are proposed.

An additional electroweak penguin contribution occurs whenever one has $M_1
M_8$ or $M_1 M_1$ production.  Since it always appears in a fixed combination
with respect to $P_1$, we have one further amplitude
\beq
p_1 \equiv P_1 - \frac{1}{3}P_{EW}~~~,
\eeq
in addition to (6)--(8), describing decays involving singlets.
\bigskip

\begin{figure}
\centerline{\epsfysize = 1.8in \epsffile{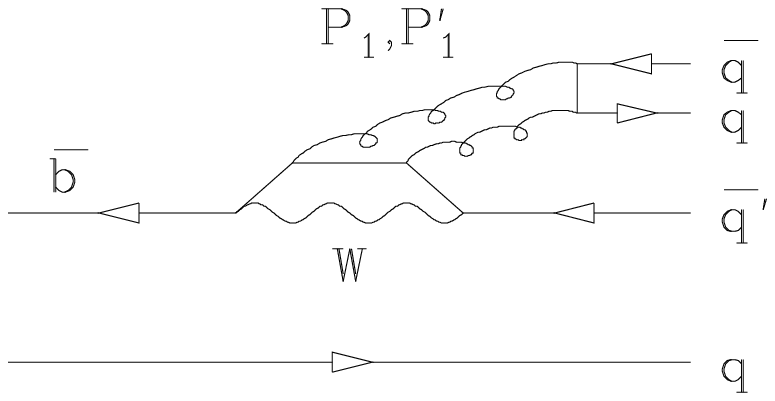}}
\caption{Graph contributing to $B \to M_1 M_8$ and $B \to M_1 M_1$.  The coiled
lines denote a color singlet exchange due to two or more gluons.}
\end{figure}

\leftline{\bf C.  Summary of amplitudes}
\bigskip

We shall denote amplitudes corresponding to $\Delta S = 0$ without primes and
those for $|\Delta S| = 1$ with primes.  The amplitudes of interest for
$|\Delta S| = 1$ decays are expressed in terms of the four independent
contributions $t',~c',~p'$, and $p_1'$ in Table I.  The amplitudes for $B^+ \to
\eta K^+$ and $B^+ \to \eta' K^+$ for arbitrary mixing angles $\theta$ may be
obtained using the definitions (\ref{eqn:etadefs}).

\renewcommand{\arraystretch}{1.3}
\begin{table}
\caption{Decomposition of $B^+ \to MM$ amplitudes for $|\Delta S| = 1$
transitions in terms of four independent quantities.  Here $\eta_p$ and
$\eta'_p$ denote the mixtures (\ref{eqn:mix}).}
\begin{center}
\begin{tabular}{c|c c c c} \hline
Final & \multicolumn{4}{c}{Coefficient of} \\
state        &   $t'$   &   $c'$   &   $p'$    &   $p_1'$  \\ \hline
$\pi^+ K^0$  &    0     &    0     &    1      &     0     \\
$\pi^0 K^+$  & $-1/\s$  & $-1/\s$  & $-1/\s$   &     0     \\
$\eta_8 K^+$ & $-1/\sx$ & $-1/\sx$ & $1/\sx$   &     0     \\
$\eta_1 K^+$ & $1/\st$  & $1/\st$  & $2/\st$   &   $\st$   \\
$\eta_p K^+$ & $-1/\st$ & $-1/\st$ &    0      & $-1/\st$  \\
$\eta'_p K^+$ & $1/\sx$ & $1/\sx$  & $3/\sx$ & $2\sqrt{2/3}$ \\ \hline
\end{tabular}
\end{center}
\end{table}

\newpage
\centerline{\bf III. QUADRANGLE CONSTRUCTION}
\bigskip

\leftline{\bf A.  A specific example of $\eta-\eta'$ mixing}
\bigskip

The amplitudes for $B^+ \to \pi^+ K^0$ contain only a contribution from $p'$.
Since both the gluonic and electroweak penguin contributions of the amplitudes
are dominated by the top quark, the weak phase of $p'$ is $\pi = {\rm
Arg}(V_{ts} V_{tb}^*)$, which does not change sign under charge conjugation.
The same is true for the term $p_1'$.  Hence we shall seek two linear
combinations of amplitudes expressed in terms of $p'$ and $p_1'$.  When
combined with $A(B^+ \to \pi^+ K^0)$, these will form an amplitude triangle
whose shape will not change under charge conjugation.

The method can be illustrated using the special mixtures $\eta_p$ and $\eta'_p$
defined in Eq.~(\ref{eqn:mix}), which are probably close to the physical
states.  We find
\beq
\st A(B^+ \to \eta_p K^+) + \sx A(B^+ \to \eta'_p K^+) = 3(p'+p_1')~~~,
\eeq
as well as
\beq
-\s A(B^+ \to \pi^0 K^+) + \st A(B^+ \to \eta_p K^+) = p' - p_1'~~~.
\eeq
Combining these results with
\beq
A(B^+ \to \pi^+ K^0) = p'~~~,
\eeq
we form a triangle with sides $p',~p'-p_1'$, and $p' + p_1'$.  This triangle
will not change shape under charge conjugation.

For simplicity we define
$$
a(\pi^+) \equiv A(B^+ \to \pi^+ K^0)~~~,~~
a(\pi^0) \equiv -\s A(B^+ \to \pi^0 K^+)~~~,
$$
\beq \label{eqn:norm}
a(\eta_p) \equiv \st A(B^+ \to \eta_p K^+)~~~,~~
a(\eta'_p) \equiv \sx A(B^+ \to \eta'_p K^+)~~~,
\eeq
so that (10)--(12) may be transcribed as
$$
\frac{1}{3} \left[ a(\eta_p) + a(\eta'_p) \right] = p' + p_1'~~~,
$$
$$
a(\pi^0) + a(\eta_p) = p' - p_1'~~~,
$$
\beq \label{eqn:tri}
a(\pi^+) = p'~~~,
\eeq
and hence
\beq \label{eqn:quadsimpl}
\frac{4}{3} a(\eta_p) + \frac{1}{3} a(\eta'_p) + a(\pi^0) = 2a(\pi^+)~~~.
\eeq
This quadrangle relation is illustrated in Fig.~4, along with the triangle
formed by the three combinations in (\ref{eqn:tri}).

\begin{figure}
\centerline{\epsfysize = 4in \epsffile{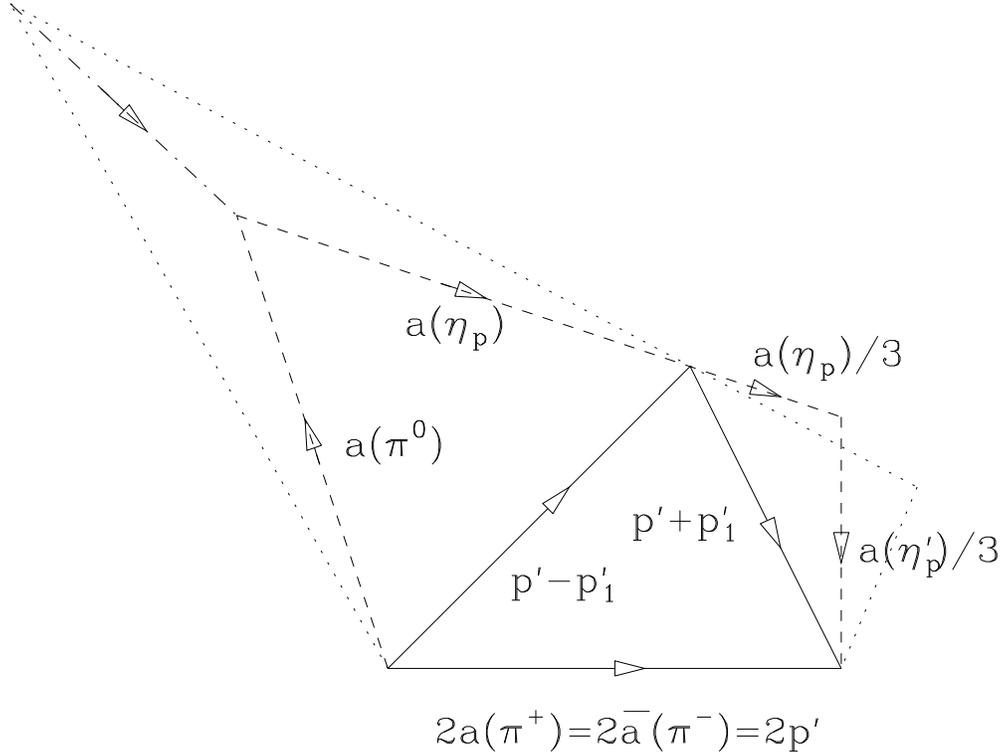}}
\caption{Quadrangle relation (\ref{eqn:quadsimpl}) satisfied by
amplitudes (\ref{eqn:norm}) for $B^+$ decays (dashed lines, with solid
line as base).  The solid triangle corresponds to the linear relation
among the combinations (\ref{eqn:tri})).  Dotted lines denote
corresponding quadrangle for $B^-$ decays.  Here $\eta_p$ and $\eta_p'$
denote the octet-singlet mixtures (\ref{eqn:mix}).  The dot-dashed
line denotes $a(\pi^0) - \bar a(\pi^0) = t' + c' - (\bar t' + \bar c')$,
whose magnitude and phase provide information on $\sin \gamma$ and a
strong phase shift difference, respectively.}
\end{figure}

The charge-conjugate processes also satisfy relations equivalent to
(\ref{eqn:tri}):

$$
\frac{1}{3} \left[ \bar a(\eta_p)+\bar a(\eta'_p) \right] = p'+p_1'~~~,
$$
$$
\bar a(\pi^0) + \bar a(\eta_p) = p' - p_1'~~~,
$$
\beq \label{eqn:tribar}
\bar a(\pi^-) = p'~~~,
\eeq
where
$$
\bar a(\pi^-) \equiv A(B^- \to \pi^- \bar K^0)~~~,~~
\bar a(\pi^0) \equiv -\s A(B^- \to \pi^0 K^-)~~~,
$$
\beq
\bar a(\eta_p) \equiv \st A(B^- \to \eta_p K^-)~~~,~~
\bar a(\eta'_p) \equiv \sx A(B^- \to \eta'_p K^-)~~~,
\eeq
and hence
\beq \label{eqn:quadsimplc}
\frac{4}{3} \bar a(\eta_p) + \frac{1}{3} \bar a(\eta'_p)
+ \bar a(\pi^0) = 2 \bar a(\pi^-)~~~.
\eeq

The triangles formed by the combinations (\ref{eqn:tri}) and
(\ref{eqn:tribar}) are identical.
Thus, the two quadrangles (\ref{eqn:quadsimpl}) and (\ref{eqn:quadsimplc})
must intersect at a point 3/4 of
the distance from their upper left-hand vertices to their upper right-hand
vertices.  The shapes of the quadrangles are thus determined, up to
discrete ambiguities which we shall discuss in Sec.~III D and in the
Appendix.  This construction
is reminiscent of one applied earlier to the decays $B \to \pi K$ and
$\bar B \to \pi \bar K$ \cite{NQ} in order to specify the shapes
of amplitude quadrangles based on isospin.

Once the quadrangles are rigid, we can form the difference
\beq \label{eqn:diff}
a(\pi^0) - \bar a(\pi^0) = t' + c' - (\bar t' + \bar c')~~~,
\eeq
where the bar denotes quantities appropriate to $B^-$ decays.  By an
argument presented earlier \cite{EWP}, this difference can be utilized
in conjunction with the amplitude for $B^\pm \to \pi^\pm \pi^0$ to
extract both a strong phase
difference and $\sin \gamma$.  We shall recapitulate this argument in
Sec.~III C.
\bigskip

\leftline{\bf B.  General mixing angle}
\bigskip

The quadrangle relations are not much more complicated for a general
mixing angle $\theta$.  We assume $\theta$ is measured by other means
(see also Sec.~IV B).  The combinations corresponding to (\ref{eqn:tri})
are
$$
\frac{\cos(\theta - \theta_0)}{\st} a(\eta)
+ \frac{\sin(\theta - \theta_0)}{\sx} a(\eta') = p' + p_1'~~~,
$$
$$
\frac{a(\eta)}{\st} + \frac{\sin(\theta - \theta_0)}{\s} a(\pi^0)
= p' \cos(\theta - \theta_0) - \st p_1' \sin \theta~~~,
$$
\beq \label{eqn:combs}
a(\pi^+) = p'~~~.
\eeq
Here $\theta_0 \equiv - {\rm arcsin}(1/\st)$, the mixing angle for which
$\eta$ would be pure strange and $\eta'$ would be pure nonstrange.  We
have retained the normalizations (\ref{eqn:norm}) for $a(\eta)$ and
$a(\eta')$.

The quadrangle relation may be written as
\beq \label{eqn:quad}
a(\eta) \left[ \sqrt{\frac{2}{3}} \csc (\theta - \theta_0)+\s \sin \theta
\cot(\theta - \theta_0) \right] + a(\eta') \sin \theta
+ a(\pi^0) = 2a(\pi^+)~~~,
\eeq
as illustrated in Fig.~5.  The two upper sides of the triangle
invariant under charge conjugation are
$$
u = \s [p' \cot(\theta-\theta_0) - \st p_1' \sin \theta \csc (\theta
- \theta_0)]~~~,
$$
\beq
v = \sx \sin \theta \csc (\theta-\theta_0) (p' + p_1')~~~.
\eeq
We have shown the two coefficients of $a(\eta)$ in (\ref{eqn:quad})
separately in order to illustrate the construction of the invariant
triangle.  The two terms in the square bracket may be combined, leading
to
\beq \label{eqn:quada}
\s a(\eta)\cos \theta + a(\eta') \sin \theta + a(\pi^0) = 2 a(\pi^+)~~~,
\eeq
which is just (\ref{eqn:firstquad}).
The combination of $\eta$ and $\eta'$ decay amplitudes appearing in
(\ref{eqn:quada}) is that corresponding to $\eta_8$, as also
pointed out in Ref.~\cite{Desh}.

Some special cases of (\ref{eqn:combs}), (\ref{eqn:quada}), and Fig.~5
may be noted.

\begin{figure}
\centerline{\epsfysize = 3in \epsffile{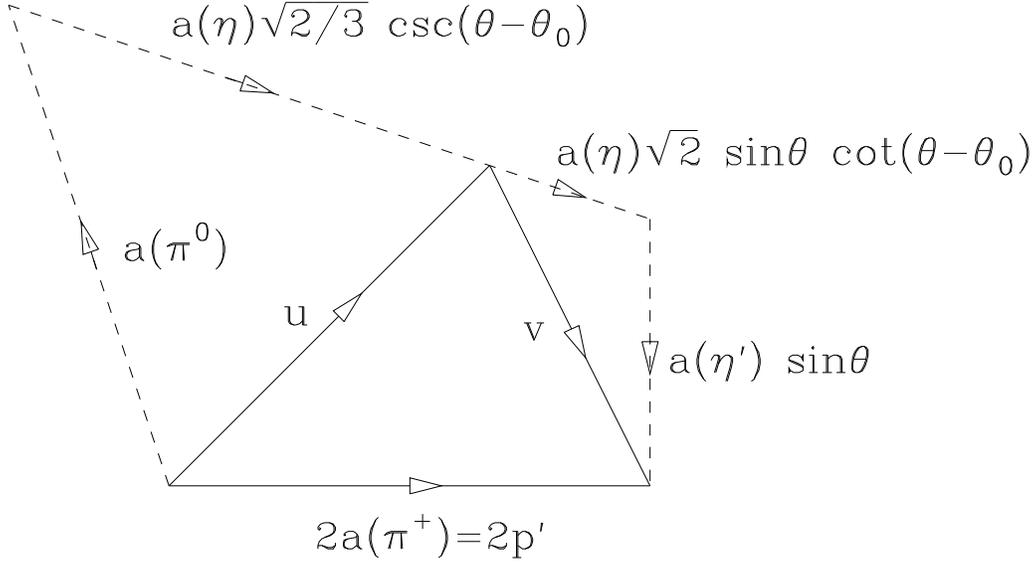}}
\caption{Quadrangle relation (\ref{eqn:quad}) satisfied by decay amplitudes
(dashed lines, with solid line as base).  The solid line corresponds to the
linear relation among the combinations (\ref{eqn:combs}).  The sum of the two
$a(\eta)$ coefficients is $\protect \sqrt{2} \cos \theta$.}
\end{figure}

When $\theta = 0$, the first two of Eqs.~(\ref{eqn:combs}) reduce to
$$
\frac{\s}{3}a(\eta_8) + \frac{1}{3\s} a(\eta_1) = p' + p_1'~~~,
$$
\beq
\frac{1}{\st} a(\eta_8) + \frac{1}{\sx} a(\pi^0) = \sqrt{\frac{2}{3}}
 p'~~~.
\eeq
The second equation may be combined with the last of (\ref{eqn:combs})
to obtain (\ref{eqn:DH}).  This relation also follows directly from
(\ref{eqn:quada}) in the limit $\theta \to 0$.

When $\theta = \theta_p$, close to the physical value, we have
$\cos(\theta - \theta_0) = \sqrt{1/3},~\sin(\theta - \theta_0) =
\sqrt{2/3}$, so the combinations (\ref{eqn:combs}) reduce to
(\ref{eqn:tri}), while (\ref{eqn:quada}) reduces to (\ref{eqn:quadsimpl}).
The sides $u$ and $v$ of the triangle in Fig.~5 reduce to $u = p' - p_1'$
and $v = p' + p_1'$.

The construction of rigid quadrangles proceeds as in the example of
Sec.~III A.  The $\bar a$ amplitudes obey a quadrangle relation similar to
(\ref{eqn:quad}) or (\ref{eqn:quada}), with the same invariant triangle.
Knowing $\theta$, we can mark off a point a suitable distance along the
$a(\eta)$ or $\bar a(\eta)$ side of each quadrangle, which must be common
to the two triangles.  As in Fig.~4, we can then determine the quantity
(\ref{eqn:diff}).

An alternative construction is clearly possible in which an invariant
triangle is constructed with one side composed of a linear combination
of $a(\pi^0)$ and $a(\eta')$ instead of $a(\pi^0)$ and $a(\eta)$.
\bigskip

\leftline{\bf C.  Determination of $\gamma$}
\bigskip

The amplitudes $a(\pi^0)$ and $\bar a(\pi^0)$, which consist of terms
containing tree and penguin weak phases, can be expressed as
$$
a(\pi^0) = |a^T_{\pi K}| e^{i \gamma} e^{i \tilde \delta_T}
- |a^P_{\pi K}| e^{i \tilde \delta_P}~~~,
$$
\beq
\bar a(\pi^0) = |a^T_{\pi K}| e^{- i \gamma} e^{i \tilde \delta_T}
- |a^P_{\pi K}| e^{i \tilde \delta_P}~~~,
\eeq
where all strong phases are
written relative to that of $p'$.  Here we have used the fact that the
weak phase of the electroweak penguin is $\pi$.  Taking the difference, we
find
\beq \label{eqn:diffg}
a(\pi^0) - \bar a(\pi^0) = 2 i \sin \gamma e^{i \tilde \delta_T}
|a^T_{\pi K}| ~~~.
\eeq

We now use flavor SU(3) to relate $|a^T_{\pi K}|$, corresponding to an
$I = 3/2~\pi K$ amplitude \cite{EWP}, to the corresponding $I = 2$
$\pi \pi$ amplitude.  [Both belong purely to the ${\bf 27}$ of SU(3).]
We find \cite{BPP,PRL,EWP}, since $\s A(B^+ \to \pi^+ \pi^0) = - (T + C) -
(P_{EW}^C + P_{EW})$
and
\beq
T'/T = C'/C = |V_{us}/V_{ud}|(f_K/f_\pi)~~~,
\eeq
(neglecting the electroweak penguin contributions) that
\beq
|a^T_{\pi K}| = |V_{us}/V_{ud}|(f_K/f_\pi) \s |A(B^\pm \to \pi^\pm
 \pi^0)|~~~.
\eeq
It does not matter whether we use the amplitude for $B^+ \to \pi^+
\pi^0$ or $B^- \to \pi^- \pi^0$; negligible CP asymmetry is expected in
the rates since electroweak penguin contributions should be small
here \cite{DH,DHT,EWP}.
\bigskip

\leftline{\bf D.  Discrete ambiguities}
\bigskip

In general, one expects $|p_1'/p'| \ne 0$ and Arg$(p_1'/p') \ne 0$, so that
there is a non-trivial invariant triangle.  In that case, it will be very
hard to avoid CP violation in rates for
$B^\pm$ decays to at least one of the modes $\pi^0 K^\pm,~\eta K^\pm$, or
$\eta' K^\pm$ if the standard picture with $\gamma \ne 0$ is correct.
We shall argue in Sec. IV C that the $\eta K^\pm$ modes
may hold the best prospect for such an asymmetry.

The two quadrangles can be degenerate,
for example, if all four sides were equal in pairs, which
would correspond to the absence of CP-violating asymmetries in decay
rates.   A limiting case is illustrated in Fig.~6.
The strong phases of $p'$ and $p_1'$ are identical in this
example; the invariant triangle has zero area.

The folded quadrangles in Fig.~6 illustrate a discrete
ambiguity which will hold in general.  The intersection of suitably
chosen points along the $a(\eta)$ and $\bar a(\eta)$ sides of the
quadrangles may be constructed either with the quadrangles as shown
in Fig.~5, or folded as in Fig.~6.  When CP violation is not present in
rates, so that the sides of the quadrangles are equal in pairs, one
solution will correspond to $\sin \gamma = 0$, while the other (as
illustrated in Fig.~6) will give a value of $\sin \gamma$ which may be
compared with the range inferred from analysis of CKM parameters
based on other experiments (such as CP violation in neutral kaon decays).
Solutions in which the difference (\ref{eqn:diffg}) yields an unphysical
value $|\sin \gamma| > 1$ may be rejected.

\begin{figure}
\centerline{\epsfysize = 1.8in \epsffile{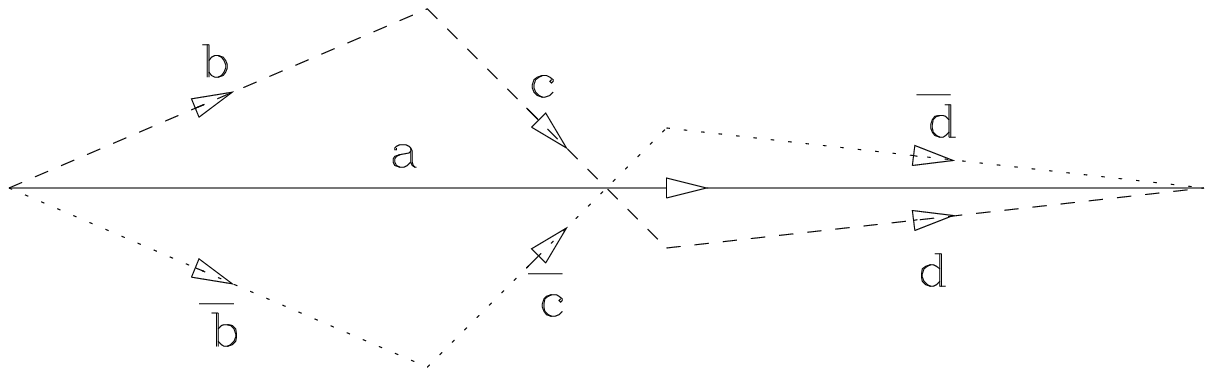}}
\caption{Example of degenerate quadrangles with no CP violation in $B^\pm$
decay modes.  Here $a \equiv 2 a(\pi^+) = 2 \bar a(\pi^-)$, $b \equiv
a(\pi^0)$, $c \equiv a(\eta) \protect \sqrt{2} \cos \theta$, $d \equiv a(\eta')
\sin \theta$, with barred quantities referring to corresponding $B^-$ decay
amplitudes.}
\end{figure}

Another trivial ambiguity corresponds to flipping Figs.~4 and
5 about the horizontal axis.  By reference
to Eq.~(\ref{eqn:diffg}), one can see that this corresponds merely to
the replacement $\tilde \delta_T \to \pi - \tilde \delta_T$, but no
change in $\gamma$.

The most general class of discrete ambiguities is illustrated by the
solution presented in the Appendix.  There, one works backward from the
known magnitudes of $a(\eta)$ and $\bar a(\eta)$ and assumes a variable
relative phase $\phi$ between them.  Then, for given $|a(\pi^0)|$,
$|\bar a(\pi^0)|$, $|a(\eta')|$, and $|\bar a(\eta')|$,
one finds a four-fold ambiguity in the solution for $|a(\pi^+)| =
|\bar a(\pi^-)|$.  The correct values of $\phi$ are those for which
$|a(\pi^+)|$ equals the observed value.
\bigskip

\centerline{\bf IV.  PRACTICAL CONSIDERATIONS}
\bigskip

\leftline{\bf A. SU(3) breaking}
\bigskip

The decays in question do not all have the same phase space.  The
correction factor $(\sim p_{\rm cm}$) is expected to be relatively
small, since
$p_{\rm cm}^{\pi K} = 2.61$ MeV/$c$,
$p_{\rm cm}^{\eta K} = 2.59$ MeV/$c$, and
$p_{\rm cm}^{\eta' K} = 2.53$ MeV/$c$.

More important is the violation of SU(3) associated with the difference
between creation of nonstrange and strange quark pairs in the final
state in penguin ($P'$) amplitudes.  As discussed in a previous
analysis \cite{SU3}, one does not really have a way to estimate this
term.  The creation of strange quark pairs occurs in the decays $B^+ \to
\eta K^+$ and $B^+ \to \eta' K^+$, but not in $B^+ \to \pi K$ decays.
For the mixed $\eta = \eta_p$ with $\theta = \theta_p = 19.5^{\circ}$,
exact SU(3) symmetry between nonstrange and strange quark pair
creation would lead to the vanishing of the $p'$  contribution
(see Table I), which one might expect \cite{EWP,SU3} to be dominant.
Assuming $|p_1'| < |p'|$, a
suppression of the rate for $B^+ \to \eta K^+$ in comparison with the
$B^+ \to \pi K$ and $\eta' K^+$ modes would be one piece of
circumstantial evidence in favor of SU(3), though the contributions
of the amplitudes besides $p'$ would have to be taken into account as well.
\bigskip

\leftline{\bf B. Nature of $\eta$ and $\eta'$}
\bigskip

The representation of $\eta$ and $\eta'$ as quark-antiquark states with
an octet-singlet mixing angle $\theta \simeq \theta_p$ is consistent
with present data \cite{GK}.  However, the admixture of non-$q \bar q$
states in the $\eta$ and $\eta'$ is less certain.  By studying the
decays $\rho \to \eta \gamma$ and $\phi \to \eta \gamma$, one can
conclude that the non-$q \bar q$ admixture of the $\eta$ is rather
small \cite{JReta}.  The $u \bar u + d \bar d$ content of the $\eta'$ is
probed by the decay $\eta' \to \rho \gamma$, but one must await the
measurement of the rate for $\phi \to \eta' \gamma$ to learn the $s \bar
s$ content of the $\eta'$ \cite{JReta}.  We have assumed that the
gluonic content of $\eta'$ is small.
\bigskip

\leftline{\bf C.  Detection of final states}
\bigskip

It is likely that the branching ratio for $B^0 \to \pi^- K^+$ is
about $10^{-5}$ \cite{Kpisep}.  This process corresponds to an
amplitude $A(B^0 \to \pi^- K^+) = - (t' + p')$, which should be
dominated by the $p'$ contribution.  Thus $A(B^+ \to \pi^0 K^+) = -
(t' + c' + p')/\s$, also expected to be dominated by $p'$, should
correspond to a branching ratio of about (1/2) $ \times 10^{-5}$,
while $A(B^+ \to \pi^+ K^0)=p'$ should correspond to a branching ratio
of $10^{-5}$.  With a branching ratio $B(K^0 \to \pi^+ \pi^-)
\simeq 1/3$, the effective branching ratio for detecting $B^+ \to
\pi^+ K^0$ via charged particles is $(1/3) \times 10^{-5}$.

The rate for the $\eta K^+$ final state is harder to estimate.  As we
have mentioned, the $p'$ contribution to this decay vanishes in the
limit of exact SU(3) for a mixing angle of $\theta = \theta_p$, so
that $A(B^+ \to \eta_p K^+) = -(t'+c'+p_1')/\st$.  We
estimated $|c'| \simeq |t'| \simeq (1/5)|p'|$ in Ref.~\cite{EWP}, so that
a branching ratio below $10^{-6}$ is a distinct possibility.
(However, if the three terms $t'$, $c'$, and $p_1'$ are of comparable
magnitude and add constructively, the suppression of the rate for
$B^+ \to \eta K^+$ could be relatively mild.)  One must also take into
account the efficiency whereby $\eta$'s
can be reconstructed, e.g. via their decays $\eta \to \gamma \gamma$
and $\eta \to \pi^+ \pi^- \pi^0$.

One advantage of the $B^+ \to \eta K^+$ mode
is that no one amplitude need be dominant, and there is room for
differences in final-state phases (especially in comparing singlet
and non-singlet amplitudes), so that this could well be a mode in
which CP violation shows up as a difference in the branching ratios
for $B^+ \to \eta K^+$ and $B^- \to \eta K^-$.

The $\eta' K^+$ amplitude has a coefficient of $p'$ equal to $3/\sx$
for $\theta = \theta_p$.  Thus the branching ratio $B(B^+ \to \eta'
K^+)$ could well exceed $10^{-5}$.  The $\eta'$ would be detectable,
for example, through its $\rho \gamma$ and $\eta \pi^+ \pi^-$ modes.
\bigskip

\centerline{\bf V. CONCLUSIONS}
\bigskip

We have shown that one can measure the weak phase $\gamma \equiv
{\rm Arg}(V_{ub}^* V_{cb}/V_{us}^* V_{cs})$ by determining the
relative rates for $B^+ \to \pi^0 K^+$, $B^+ \to \pi^+ K^0$,
$B^+ \to \eta K^+$, $B^+ \to \eta' K^+$, and the corresponding
charge-conjugate processes.  The method is based on construction of
amplitude quadrangles satisfying a constraint that specifies their
shapes up to possible discrete ambiguities.  The difference between
the complex amplitudes for $B^+ \to \pi^0 K^+$ and $B^- \to \pi^0 K^-$,
when compared with the magnitude of the amplitude for $B^{\pm} \to
\pi^\pm \pi^0$, then yields both $\sin \gamma$ and some information on
differences of strong-interaction phase shifts.

Key assumptions in this program include the validity of nonet symmetry
for $\eta$ and $\eta'$ (testable to some extent in electromagnetic
processes \cite{Chau,JReta}), the neglect of amplitudes scaling as
$f_B$ (testable in relations proposed earlier \cite{BPP}), and the
validity of SU(3) itself (testable to a limited extent in other cases
\cite{SU3} but only indirectly here through a possible suppression of the
decays $B^\pm \to \eta K^\pm$).  Indeed, we expect the detection of
these decays to be the most demanding aspect of the present
construction.  Once the $\eta K^\pm$ mode has been seen,
the major hurdle will have been overcome, with the possibility of a
CP asymmetry in $B \to \eta K^\pm$ rates.  Along the way, we expect
the decays $B^\pm \to \eta' K^\pm$ to make a prominent appearance at the
branching ratio level of $10^{-5}$ or greater.
\bigskip

\centerline{\bf ACKNOWLEDGMENTS}
\bigskip

We thank H. Lipkin, S. Meshkov, and S. Stone for fruitful discussions, and the
Aspen Center for Physics for a congenial atmosphere in which the main part of
this collaboration was carried out. M. G. wishes to acknowledge the hospitality
of the SLAC theory group during parts of this investigation. This work was
supported in part by the United States -- Israel Binational Science Foundation
under Research Grant Agreement 94-00253/1, by the Fund for Promotion of
Research at the Technion, and by the United States Department of Energy under
Contract No. DE FG02 90ER40560.
\bigskip

\centerline{\bf APPENDIX:  EXPLICIT CONSTRUCTION OF QUADRANGLES}
\bigskip

We show in Fig.~7 the way in which the shapes of the quadrangles may be
explicitly determined using only the lengths of the sides, the fact that they
share a side, and the fact that the opposite sides intersect at a point
corresponding to a fixed ratio of their lengths.  The discussion is presented
for simplicity for the case $\theta = \theta_p = {\rm arcsin}(1/3)$.  Letters
denote magnitudes of amplitudes, whereas in Fig.~6 they denoted the complex
amplitudes.

\begin{figure}
\centerline{\epsfysize = 5.5in \epsffile{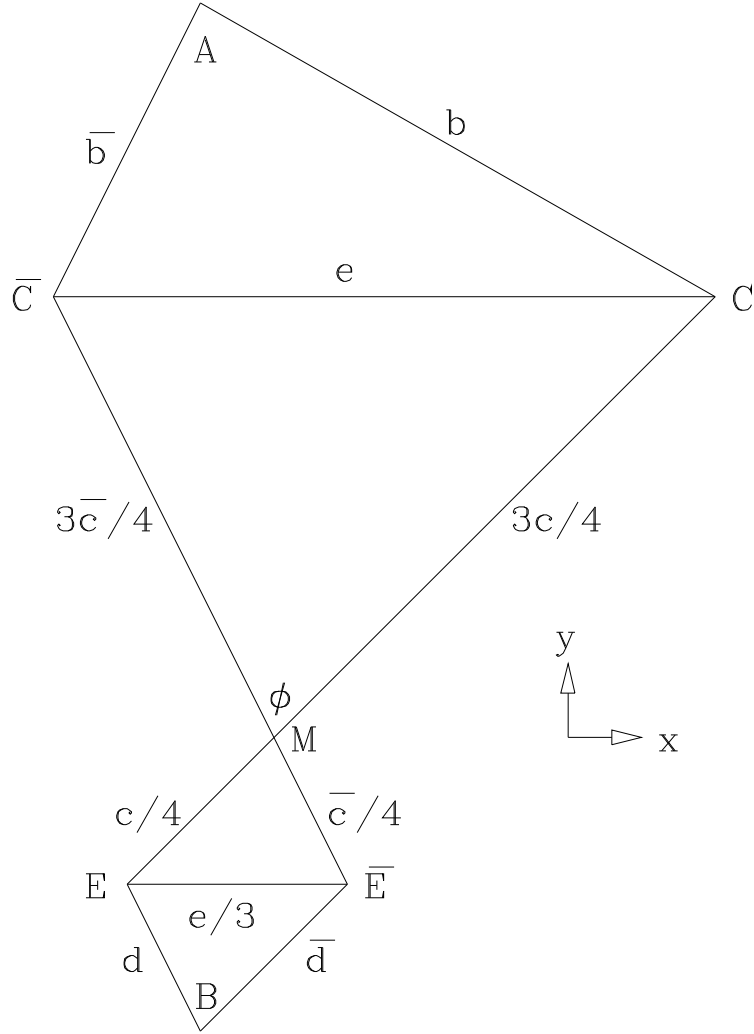}}
\caption{Explicit construction of amplitude quadrangles.  Here we denote
$|a(\pi^0)| \equiv b$, $|a(\pi^0) - \bar a(\pi^0)| \equiv e$, $|a(\eta)| \equiv
3c/4$, $|a(\eta')|/3 \equiv d$, and $2|a(\pi^+)| = 2|\bar a(\pi^-)| \equiv AB =
a$, with corresponding notation for barred quantities.}
\end{figure}

We first construct lines with length $c = (4/3)|a(\eta)|$ and $\bar c =
(4/3)|\bar a(\eta)|$ intersecting at a variable angle $\phi$ at a point 3/4 of
the distance along their lengths.  We orient these lines so that the points $C$
and $\bar C$ are joined by a line parallel to the $x$ axis, and call the
intersection point $M$ the origin.  The length of the line $C \bar C$ will be
$e = |a(\pi^0) - \bar a(\pi^0)|$; that of $E \bar E$ will be $e/3$ and $E \bar
E$ will also be parallel to the $x$ axis.

Now construct the triangle $C \bar C A$ with sides $b = |a(\pi^0)|$, $\bar b =
|\bar a(\pi^0)|$, and $e$, and the triangle $E \bar E B$ with sides $d =
|a(\eta')|/3$, $\bar d = |\bar a(\eta')|/3$, and $e/3$. The distance $a = AB$
must be given by $2|a(\pi^+)|$.  This determines the angle $\phi$. There is at
most a four-fold ambiguity in this construction, corresponding to the
possibility of flipping either triangle about its base.

It is easily seen that the above construction specifies the distance $a = AB$
up to a four-fold ambiguity, given any value of $e$ for which the two lines
$CE$ and $\bar C \bar E$ can actually intersect.  For example, the height $h_A$
of the triangle $C \bar C A$ (its projection along the $y$ axis) is
\beq
h_A = \frac{1}{2e}\left[ - \lambda(b^2, \bar b^2, e^2) \right]^{1/2}~~~,
\eeq
where $\lambda(a,b,c) \equiv a^2 + b^2 + c^2 - 2ab - 2ac - 2bc$, with similar
expressions for the heights $g_A,~g_B = g_A/3$, and $h_B$ of the respective
triangles $C \bar C M$, $E \bar E M$, and $E \bar E B$.  Then $y_A - y_B = g_A
+ g_B \pm h_A \pm h_B$ is specified up to the four-fold ambiguity mentioned
above.  The $x$ coordinates of $A$ and $B$ may be obtained from expressions
such as
\beq
x_C - x_A = \frac{1}{2e}\left[ e^2 + b^2 - \bar b^2 \right]~~~,
\eeq
with similar expressions for $x_C$, $x_E$, and $x_B - x_E$.  Thus one can also
obtain $x_A - x_B$, and hence $AB = [(x_A - x_B)^2 + (y_A - y_B)^2]^{1/2}$.

\def \ajp#1#2#3{Am.~J.~Phys.~{\bf#1}, #2 (#3)}
\def \apny#1#2#3{Ann.~Phys.~(N.Y.) {\bf#1}, #2 (#3)}
\def \app#1#2#3{Acta Phys.~Polonica {\bf#1}, #2 (#3)}
\def \arnps#1#2#3{Ann.~Rev.~Nucl.~Part.~Sci.~{\bf#1}, #2 (#3)}
\def \cmp#1#2#3{Commun.~Math.~Phys.~{\bf#1}, #2 (#3)}
\def \cmts#1#2#3{Comments on Nucl.~Part.~Phys.~{\bf#1}, #2 (#3)}
\def \cn{Collaboration}
\def \corn93{{\it Lepton and Photon Interactions:  XVI International Symposium,
Ithaca, NY August 1993}, AIP Conference Proceedings No.~302, ed.~by P. Drell
and D. Rubin (AIP, New York, 1994)}
\def \cp89{{\it CP Violation,} edited by C. Jarlskog (World Scientific,
Singapore, 1989)}
\def \dpff{{\it The Fermilab Meeting -- DPF 92} (7th Meeting of the American
Physical Society Division of Particles and Fields), 10--14 November 1992,
ed. by C. H. Albright \ite~(World Scientific, Singapore, 1993)}
\def \dpf94{DPF 94 Meeting, Albuquerque, NM, Aug.~2--6, 1994}
\def \efi{Enrico Fermi Institute Report No. EFI}
\def \el#1#2#3{Europhys.~Lett.~{\bf#1}, #2 (#3)}
\def \f79{{\it Proceedings of the 1979 International Symposium on Lepton and
Photon Interactions at High Energies,} Fermilab, August 23-29, 1979, ed.~by
T. B. W. Kirk and H. D. I. Abarbanel (Fermi National Accelerator Laboratory,
Batavia, IL, 1979}
\def \hb87{{\it Proceeding of the 1987 International Symposium on Lepton and
Photon Interactions at High Energies,} Hamburg, 1987, ed.~by W. Bartel
and R. R\"uckl (Nucl. Phys. B, Proc. Suppl., vol. 3) (North-Holland,
Amsterdam, 1988)}
\def \ib{{\it ibid.}~}
\def \ibj#1#2#3{~{\bf#1}, #2 (#3)}
\def \ichep72{{\it Proceedings of the XVI International Conference on High
Energy Physics}, Chicago and Batavia, Illinois, Sept. 6--13, 1972,
edited by J. D. Jackson, A. Roberts, and R. Donaldson (Fermilab, Batavia,
IL, 1972)}
\def \ijmpa#1#2#3{Int.~J.~Mod.~Phys.~A {\bf#1}, #2 (#3)}
\def \ite{{\it et al.}}
\def \jmp#1#2#3{J.~Math.~Phys.~{\bf#1}, #2 (#3)}
\def \jpg#1#2#3{J.~Phys.~G {\bf#1}, #2 (#3)}
\def \lkl87{{\it Selected Topics in Electroweak Interactions} (Proceedings of
the Second Lake Louise Institute on New Frontiers in Particle Physics, 15--21
February, 1987), edited by J. M. Cameron \ite~(World Scientific, Singapore,
1987)}
\def \ky85{{\it Proceedings of the International Symposium on Lepton and
Photon Interactions at High Energy,} Kyoto, Aug.~19-24, 1985, edited by M.
Konuma and K. Takahashi (Kyoto Univ., Kyoto, 1985)}
\def \mpla#1#2#3{Mod.~Phys.~Lett.~A {\bf#1}, #2 (#3)}
\def \nc#1#2#3{Nuovo Cim.~{\bf#1}, #2 (#3)}
\def \np#1#2#3{Nucl.~Phys.~{\bf#1}, #2 (#3)}
\def \pisma#1#2#3#4{Pis'ma Zh.~Eksp.~Teor.~Fiz.~{\bf#1}, #2 (#3) [JETP Lett.
{\bf#1}, #4 (#3)]}
\def \pl#1#2#3{Phys.~Lett.~{\bf#1}, #2 (#3)}
\def \plb#1#2#3{Phys.~Lett.~B {\bf#1}, #2 (#3)}
\def \pr#1#2#3{Phys.~Rev.~{\bf#1}, #2 (#3)}
\def \pra#1#2#3{Phys.~Rev.~A {\bf#1}, #2 (#3)}
\def \prd#1#2#3{Phys.~Rev.~D {\bf#1}, #2 (#3)}
\def \prl#1#2#3{Phys.~Rev.~Lett.~{\bf#1}, #2 (#3)}
\def \prp#1#2#3{Phys.~Rep.~{\bf#1}, #2 (#3)}
\def \ptp#1#2#3{Prog.~Theor.~Phys.~{\bf#1}, #2 (#3)}
\def \rmp#1#2#3{Rev.~Mod.~Phys.~{\bf#1}, #2 (#3)}
\def \rp#1{~~~~~\ldots\ldots{\rm rp~}{#1}~~~~~}
\def \si90{25th International Conference on High Energy Physics, Singapore,
Aug. 2-8, 1990}
\def \slc87{{\it Proceedings of the Salt Lake City Meeting} (Division of
Particles and Fields, American Physical Society, Salt Lake City, Utah, 1987),
ed.~by C. DeTar and J. S. Ball (World Scientific, Singapore, 1987)}
\def \slac89{{\it Proceedings of the XIVth International Symposium on
Lepton and Photon Interactions,} Stanford, California, 1989, edited by M.
Riordan (World Scientific, Singapore, 1990)}
\def \smass82{{\it Proceedings of the 1982 DPF Summer Study on Elementary
Particle Physics and Future Facilities}, Snowmass, Colorado, edited by R.
Donaldson, R. Gustafson, and F. Paige (World Scientific, Singapore, 1982)}
\def \smass90{{\it Research Directions for the Decade} (Proceedings of the
1990 Summer Study on High Energy Physics, June 25 -- July 13, Snowmass,
Colorado), edited by E. L. Berger (World Scientific, Singapore, 1992)}
\def \stone{{\it B Decays}, edited by S. Stone (World Scientific, Singapore,
1994)}
\def \tasi90{{\it Testing the Standard Model} (Proceedings of the 1990
Theoretical Advanced Study Institute in Elementary Particle Physics, Boulder,
Colorado, 3--27 June, 1990), edited by M. Cveti\v{c} and P. Langacker
(World Scientific, Singapore, 1991)}
\def \yaf#1#2#3#4{Yad.~Fiz.~{\bf#1}, #2 (#3) [Sov.~J.~Nucl.~Phys.~{\bf #1},
#4 (#3)]}
\def \zhetf#1#2#3#4#5#6{Zh.~Eksp.~Teor.~Fiz.~{\bf #1}, #2 (#3) [Sov.~Phys. -
JETP {\bf #4}, #5 (#6)]}
\def \zpc#1#2#3{Zeit.~Phys.~C {\bf#1}, #2 (#3)}

\end{document}